\documentstyle[11pt,newpasp,twoside,epsf]{article}

\def\edcomment#1{\iffalse\marginpar{\raggedright\sl#1\/}\else\relax\fi}
\reversemarginpar

\begin{document}

\title{Star Formation Rate estimators: 
[O{\footnotesize\bf\,II}]{\boldmath$\lambda$}3727 vs.\ H{\boldmath$\alpha$} for
local star-forming galaxies}

\author{A. Arag\'on-Salamanca} 
\affil{School of Physics \& Astronomy, University of Nottingham, 
Nottingham NG7 2RD, UK}
\author{A. Alonso-Herrero}
\affil{Steward Observatory, University of Arizona, Tucson, AZ 85721, USA}
\author{J. Gallego,  C.E. Garc\'{\i}a-Dab\'o, P.G. P\'erez-Gonz\'alez, 
J. Zamorano} 
\affil{Departamento de Astrof\'{\i}sica, Universidad
       Complutense de Madrid, 
       Facultad CC F\'{\i}sicas, Ciudad Universitaria, E-28040 Madrid, Spain}
\author{A. Gil de Paz}
\affil{The Observatories of the Carnegie Institution of Washington, 813 Santa
Barbara Street, Pasadena, CA 91101, USA}

\begin{abstract}

The \mbox{[O\,{\sc ii}]}$\lambda$3727 emission line is frequently used as an indicator of the
star formation rate (SFR) despite its complex dependence on metallicity and
excitation  conditions. We have analysed the properties of the \mbox{[O\,{\sc ii}]} and
H$\alpha$ emission lines for a complete sample of local H$\alpha$-selected
galaxies,   the Universidad Complutense de Madrid (UCM) survey.  We find a
large scatter in the ${\rm \mbox{[O\,{\sc ii}]}}/{{\rm H}\alpha}$ line ratios, although the
scatter in the extinction-corrected ${\rm \mbox{[O\,{\sc ii}]}}^0/{{\rm H}\alpha^0}$ ratio
is considerably smaller.  We also find that the ${\rm \mbox{[O\,{\sc ii}]}}/{{\rm H}\alpha}$
ratios are reasonably well correlated with the absolute $B$- and $K$-band
magnitudes and with $EW({\rm \mbox{[O\,{\sc ii}]}})$.  However, the extinction-corrected ${\rm
\mbox{[O\,{\sc ii}]}^0}/{{\rm H}\alpha^0}$ ratio is largely independent of these quantities,
indicating  that extinction is the main driver of the correlations.  These
correlations allow us to statistically  predict---with varying degrees of
accuracy---the observed  and extinction-corrected H$\alpha$ fluxes from the
observed \mbox{[O\,{\sc ii}]} flux using the information contained in $EW({\rm \mbox{[O\,{\sc ii}]}})$
and/or  the absolute magnitudes, but extreme caution is  needed to make sure
that the sample selection effects are correctly  taken into account.

\end{abstract}

\section{Introduction}

Measuring the evolution of the SFR density of the universe as a function of 
look-back time is essential to understand the formation and evolution of
galaxies.   The H$\alpha$ luminosity is one of the best optical estimators of
the current SFR, modulo the Initial Mass Function (e.g., Kennicutt 1992,
1998).    Unfortunately, H$\alpha$ can only be observed with optical CCDs out
to $z\approx 0.4$.  At higher redshifts, the  \mbox{[O\,{\sc ii}]}$\lambda$3727 line is
frequently used as a SFR estimator due to its availability in optical
spectroscopic samples. However, the transformation between the number of
ionising  photons  (which is a direct consequence of the current SFR)  and the
intensity or equivalent width ($EW$) of this line is 
strongly  dependent on the excitation conditions and metallicity.
Previous empirical calibrations of the SFR in  terms of the \mbox{[O\,{\sc ii}]} line (e.g.,
Gallagher, Bushouse,  \& Hunter 1989; Kennicutt 1992; Guzm\'an et al.\ 1997)
vary within a factor of a few.  Among other problems, these calibrations may
suffer from small number statistics, incompleteness in luminosities, starburst
ages and/or physical conditions of their samples, and uncertain extinction
corrections.  

We have analysed the \mbox{[O\,{\sc ii}]} emission line properties of a local  sample of
H$\alpha$-selected star forming galaxies (the UCM sample) and derived
reasonably precise  empirical calibrations between the \mbox{[O\,{\sc ii}]} and H$\alpha$
luminosities as a function of the  galaxy population properties.   In a recent
paper, Jansen, Franx \& Fabricant (2001) carried out a similar study for the
Nearby Field Galaxy Survey (NFGS, Jansen et al.\ 2000a,b)  and found, among
other things, that the observed  \mbox{[O\,{\sc ii}]}/H$\alpha$ ratio varies by a factor  of~7
at luminosities near $M_B^*$, and that the \mbox{[O\,{\sc ii}]}/H$\alpha$ ratio  is inversely
correlated with luminosity.  Our study is, in many ways, complementary of that
of Jansen et al., but  we use  an H$\alpha$-selected (thus SFR-selected) sample
of galaxies instead of a $B$-band selected one. We confirm most of the findings
of Jansen et al., but we also demonstrate that using an extra piece of
information, namely the $EW$ of \mbox{[O\,{\sc ii}]}, much of the scatter in the \mbox{[O\,{\sc ii}]}/H$\alpha$
ratio can be removed, allowing a much more accurate calibration of \mbox{[O\,{\sc ii}]} vs.\
SFR.  

Our basic dataset comes from Universidad Complutense de  Madrid (UCM) survey of
H$\alpha$ selected galaxies. Optical spectroscopy  (Gallego et al.\ 1996; 1997)
and optical and near-infrared photometry  (P\'erez-Gonz\'alez et al. 2000, 
2002; Gil de Paz et al. 2000) are available for this sample. Based on their
spectroscopic properties, the  UCM galaxies have been roughly divided into two
main classes:  H\,{\sc ii}-like and disk-like  galaxies (see Gil de Paz et al.
2000). We will compare the  properties of the UCM galaxies with those of the
NFGS galaxies of Jansen et al. A detailed description of our study can be found
in Arag\'on-Salamanca et al. (2002).  We assume $H_0 =
50\,$km$\,$s$^{-1}\,$Mpc$^{-1}$ throughout.

\section{Results:}

\begin{figure}
\plotone{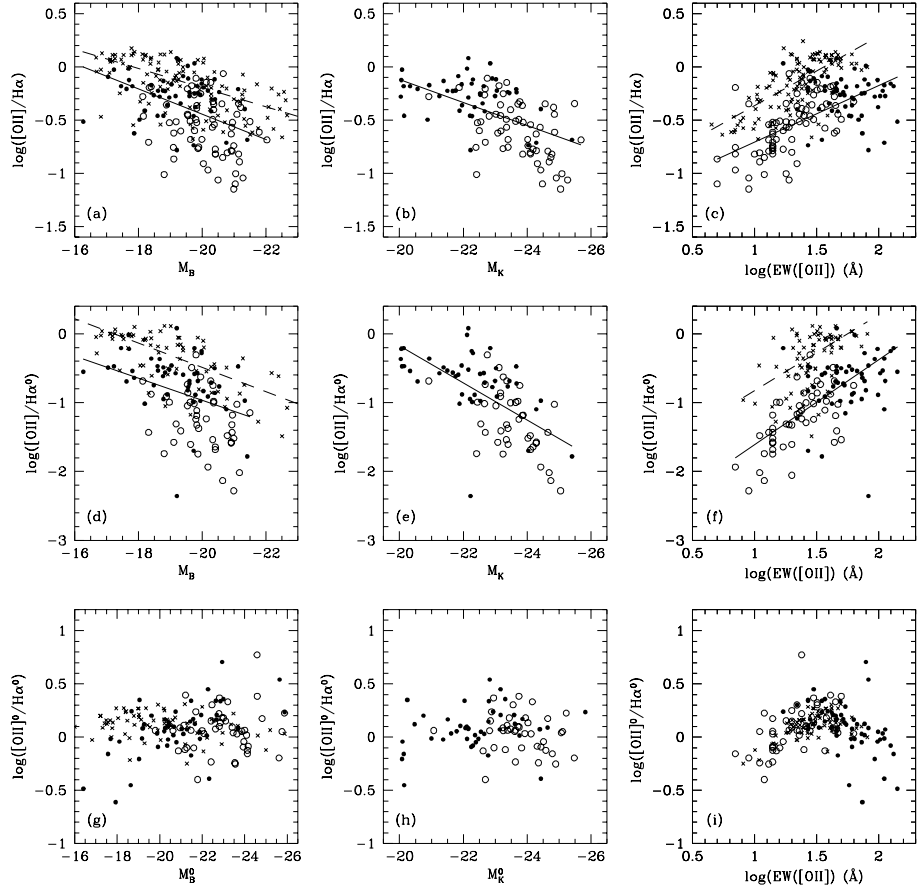}

\caption{{\bf (a)}, {\bf (b)} and {\bf (c)}  The observed $\log({\rm
\mbox{[O\,{\sc ii}]}}/{\rm H}\alpha)$ vs.\ $M_B$, $M_K$ and  $\log EW($\mbox{[O\,{\sc ii}]}$)$ for the  UCM
disk galaxies (open circles) and H{\sc ii} galaxies (small  filled circles), 
where ${\rm \mbox{[O\,{\sc ii}]}}$ and ${\rm H}\alpha$ are the observed 
emission line fluxes. The  solid line is a linear fit to the data. The crosses
and the  dashed line correspond to the NFGS sample. {\bf (d)}, {\bf (e)} and
{\bf (f)}  As before, but using the extinction-corrected H$\alpha$ flux. {\bf
(g)}, {\bf (h)} and {\bf (i)}  As before, but using extinction-corrected
H$\alpha$ and \mbox{[O\,{\sc ii}]} fluxes. } 

\end{figure}

In Figure~1 we plot the ${\rm\mbox{[O\,{\sc ii}]}}/{\rm H}\alpha$ ratio (both  observed and
corrected from extinction using the Balmer 
decrement\footnote{Throughout this paper ${\rm \mbox{[O\,{\sc ii}]}}^0$ 
and ${{\rm H}\alpha^0}$ indicate 
extinction-corrected emission-line fluxes.}) for the UCM and NFGS
galaxies as a function of absolute $B$- and $K$-band magnitudes and
$EW($\mbox{[O\,{\sc ii}]}$)$. There is a large scatter in the ${\rm \mbox{[O\,{\sc ii}]}}/{{\rm H}\alpha}$ and
${\rm \mbox{[O\,{\sc ii}]}}/{{\rm H}\alpha^0}$ ratios for all the galaxies, but the scatter 
in the extinction-corrected ${\rm \mbox{[O\,{\sc ii}]}}^0/{{\rm H}\alpha^0}$ ratio is
considerably smaller, indicating that much of this scatter is due to
differences in extinction.

Figure~1 also reveals significant difference in the median ${\rm \mbox{[O\,{\sc ii}]}}/{{\rm
H}\alpha}$ and ${\rm \mbox{[O\,{\sc ii}]}}/{{\rm H}\alpha^0}$ ratios for the UCM and NFGS
galaxies. These ratios are considerably smaller for the UCM galaxies.  However,
the  extinction-corrected ${\rm \mbox{[O\,{\sc ii}]}^0}/{{\rm H}\alpha^0}$ ratios for both
samples agree quite well, indicating that the line ratio differences between
both samples are mainly due to a difference in the mean extinction. This is 
readily explained by the fact that the NFGS sample has been selected in the
$B$-band, while the UCM sample was selected in the red (H$\alpha$), and thus it
is not surprising that galaxies with higher extinction are underrepresented in
the blue selected sample. Therefore, differences in sample selection can yield
very large differences in the line emission properties. 

The third conclusion from these plots is  that the  ${\rm \mbox{[O\,{\sc ii}]}}/{{\rm
H}\alpha}$ and ${\rm \mbox{[O\,{\sc ii}]}}/{{\rm H}\alpha^0}$ ratios for disk-like and
H\,{\sc ii}-like galaxies are very different.  However,  the average  ${\rm
\mbox{[O\,{\sc ii}]}^0}/{{\rm H}\alpha^0}$ ratio is the same for both classes of galaxies.
Again, the mixture of galaxy types in a given sample can have large effects in
the line ratios. 

\begin{figure}
\plotone{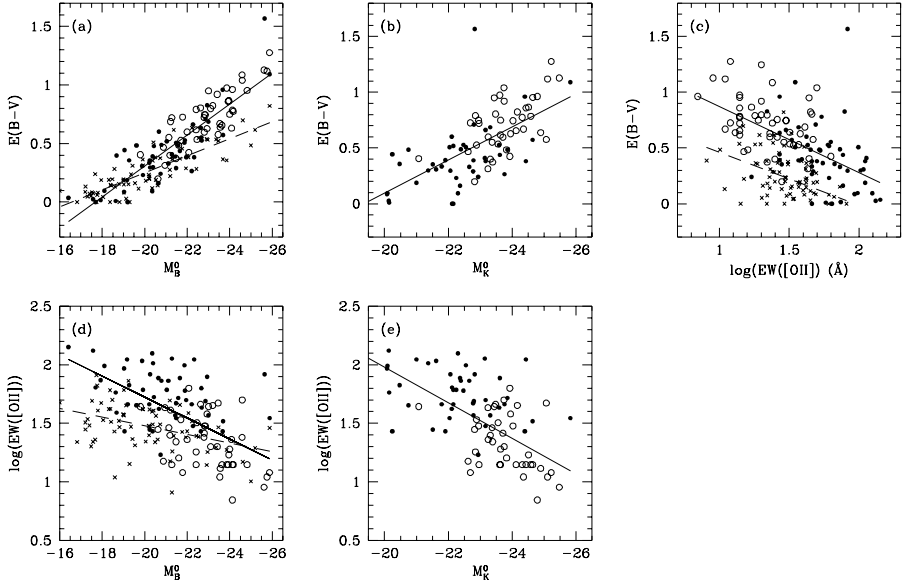}

\vspace*{-4.5cm}

\caption{{\bf (a)} The Balmer-decrement derived color excesses vs.\ the
extinction-corrected absolute magnitudes  in the $B$-band for the UCM and NFGS
galaxies. Symbols as in Figure~1. {\bf (b)} as (a) but vs.\ $M_K^0$. {\bf (c)}
The derived colour excesses vs.\ $\log EW($\mbox{[O\,{\sc ii}]}$)$. {\bf (d)} and {\bf (e)}
The measured $EW($\mbox{[O\,{\sc ii}]}$)$ vs.\ the absolute magnitudes $M_B^0$ and $M_K^0$
respectively.  The solid and dashed  lines show linear fits to the observed
trends for the UCM and NFGS galaxies respectively. } 

\end{figure}

We also find that, although with significant scatter,  the ${\rm \mbox{[O\,{\sc ii}]}}/{{\rm
H}\alpha}$ and ${\rm \mbox{[O\,{\sc ii}]}}/{{\rm H}\alpha^0}$ ratios are reasonably well
correlated with the absolute $B$- and $K$-band magnitudes and with $EW({\rm
\mbox{[O\,{\sc ii}]}})$. However, the extinction-corrected ${\rm \mbox{[O\,{\sc ii}]}^0}/{{\rm H}\alpha^0}$
ratio is largely independent of these quantities, indicating  that extinction
is the main driver of these correlations.  
This is supported by Figure~2, which shows quite reasonable 
correlations between the extinction and the absolute magnitudes of the galaxies.
The extinction is also correlated with $EW({\rm \mbox{[O\,{\sc ii}]}})$. This figure also
shows that the typical extinction for the NFGS galaxies is significantly
smaller than for the UCM galaxies. 

These correlations could allow us to statistically predict---with varying
degrees of accuracy---the observed  and even  extinction-corrected H$\alpha$
fluxes from the observed \mbox{[O\,{\sc ii}]} flux using the information contained in the
$EW({\rm \mbox{[O\,{\sc ii}]}})$ and/or  the absolute magnitudes, provided that the sample
selection effects are correctly taken into account. Indeed, it is very
important to stress that these correlations apply, strictly speaking, to
samples with similar properties to the ones presented here. If the selection of
the sample is different, significant systematics could be present, as shown by
the differences in behavior between the UCM and the NFGS galaxies. Moreover,
there is no guarantee that high redshift galaxy samples would not show
significant evolution in their line ratios (e.g., due to chemical evolution)
which could, in principle, invalidate any locally-derived calibration.

Any calibration derived from these empirical correlations
would have a significant intrinsic scatter, and
thus the predictions for individual galaxies will be quite uncertain.  
However, if one is only interested in statistical properties of large  samples
(e.g., determining the  star-formation rate density of large regions of the
Universe), the problem of the large scatter would be alleviated. But it is
crucial to take into account the properties of the sample, in particular  the
selection criteria, luminosity range and mixture of galaxy types, since
important systematic effects  are also present on top of the large scatter.
Obviously, just using large samples would not solve the  systematic problems.

Finally, there is another problem we have not discussed and needs to be
considered: the effect of galaxies with no \mbox{[O\,{\sc ii}]} emission on the statistical
determination of SFRs and SFR densities. A significant fraction of the UCM
galaxies (20\%) have H$\alpha$ emission but no  detectable \mbox{[O\,{\sc ii}]} emission
($EW({\rm \mbox{[O\,{\sc ii}]}}>-5\,$\AA). This means that by  only counting the \mbox{[O\,{\sc ii}]} 
emitting UCM galaxies, we would miss $\sim15$\% of the  extinction-corrected
H$\alpha$ flux, i.e., $\sim15$\% of the SFR. This problem  becomes negligible
for a blue-selected sample such as the NFGS since, due mainly to its lower
average extinction, virtually all NFGS galaxies with H$\alpha$ emission have
detectable \mbox{[O\,{\sc ii}]}.


\acknowledgments This research was supported in part by the Spanish Programa
Nacional de Astronom\'\i a y Astrof\'\i sica. AAS acknowledges generous
financial support from the Royal Society.   AGdP acknowledges financial support
from NASA. PGPG acknowledges  financial support from the Spanish Ministerio de
Educaci\'on y Cultura. CEGD acknowledges financial support from the UCM.


\section*{Discussion}

\noindent
{\it Kennicutt:} I was surprised by the large shift in 
${\rm \mbox{[O\,{\sc ii}]}}/{{\rm H}\alpha}$ ratio between your UCM disk
galaxy sample and the NFGS objects because my impression was that the types of
galaxies were similar (despite the different selection methods). Could part of
the difference arise because your spectra sample only the central parts of the
galaxies? \\

\noindent {\it Arag\'on-Salamanca:} I don't think so. The UCM spectra were
taken with slit widths that approximately match the size of the galaxies (at
the expense of spectral resolution). Thus the spectra used here are close to
``global'' spectra. Note that the majority of the UCM galaxies are
significantly more distant than the NFGS galaxies, and therefore have much
smaller angular sizes. Moreover, the fact that the extinction-corrected ${\rm
\mbox{[O\,{\sc ii}]}^0}/{{\rm H}\alpha^0}$ ratios of the UCM and NFGS galaxies
match very well gives us confidence that  extinction is the main driver of the
observed differences, and that the intrinsic ratios are indeed the same. \\

\noindent
{\it R.\ Terlevich:} In a recent analysis I did with Elena Terlevich and 
Daniel Rosa we found that using the 
${{\rm H}\alpha}/{{\rm H}\beta}$ ratio without taking into account the
the underlying Balmer absorption you tend to overestimate the extinction.
Could this be related to the relation found between the   
$EW({\rm \mbox{[O\,{\sc ii}]}})$ and extinction? \\

\noindent  {\it Arag\'on-Salamanca:} It is true that the observed ${{\rm
H}\alpha}/{{\rm H}\beta}$ ratio can be affected by stellar absorption  and thus
bias the extinction measurements. But the NFGS Balmer decrement  measurements
were corrected for stellar absorption, and the trend between  $EW({\rm
\mbox{[O\,{\sc ii}]}})$ and extinction is similar to the one found for the UCM
galaxies (albeit, with higher average extinction, as discussed above; cf.\
Figure~2c). Moreover, being aware of this problem,  we only attempted to
determine Balmer-decrement extinctions for UCM  galaxies with reasonably strong
emission lines ($EW({\rm H}\beta)<-5\,$\AA), where this effect should be
smaller. It is interesting to note that  $EW({\rm \mbox{[O\,{\sc ii}]}})$
anti-correlates with extinction {\it and\/} luminosity (Figure~2). Because of
the luminosity-metallicity relation, it is not unreasonable to expect
metal-rich galaxies to have more dust and thus more extinction.   \\



\begin{references}

\reference Arag\'on-Salamanca, A.,  Alonso-Herrero, A.,  Gallego, J., 
Garc\'{\i}a-Dab\'o, C.E.,  Gil de Paz, A.,  P\'erez-Gonz\'alez, P.G.   \&
Zamorano, J. 2002, \aj, submitted

\reference Gallagher, J.S., Bushouse, H. \& Hunter, D.A. 1989,  \aj, 97, 700

\reference Gallego J., Zamorano J., Alonso O. \&  
Vitores A.G.  1996, \aaps, 120, 323

\reference Gallego J., Zamorano J., Rego M. \& Vitores A.G. 1997, \apj, 
475, 502

\reference  Gil de Paz, A., Arag\'on-Salamanca, A., Gallego, J.,
Alonso-Herrero, A.,  Zamorano, J. \& Kauffmann, G. 2000, \mnras, 316, 357

\reference Guzm\'an, R., Gallego, J., Koo, D.C., Phillips, A.C., Lowenthal,
J.D., Faber, S.M., Illingworth, G.D. \& Vogt, N.P. 1997, \apj, 489, 559

\reference Jansen, R.A., Franx, M., Fabricant, D.G. \& Caldwell, N. 2000a,
\apjs, 126, 271

\reference Jansen, R.A., Fabricant,  D.G., Franx, M. \& Caldwell, N. 2000b,
\apjs, 126, 331

\reference Jansen, R.A., Franx, M. \& Fabricant, D. 2001, \apj, 551, 825

\reference Kennicutt, R.C. 1992, \apj, 388, 310

\reference Kennicutt, R.C. 1998, \araa, 36, 189

\reference Per\'ez-Gonz\'alez, P.G., Zamorano, J., Gallego, J. \& Gil de Paz,
A. 2000, \aaps, 141, 409


\reference Per\'ez-Gonz\'alez, P.G.,  Gil de Paz, A.,  Zamorano, J.,  Gallego,
J., Alonso-Herrero, A. \& Arag\'on-Salamanca, A. 2002, \mnras, in press, 
astro-ph/0209396

\end{references}
\end{document}